\documentclass[aps,prd,onecolumn,superscriptaddress,nofootinbib,showpacs]{revtex4}

\usepackage{graphicx}
\usepackage{slashed}
\usepackage{amsmath,bbm,latexsym,amssymb}
\usepackage{epsfig}
\usepackage{epstopdf}

\newcommand{\be}{\begin{equation}}
\newcommand{\ee}{\end{equation}}
\newcommand{\ba}{\begin{eqnarray}}
\newcommand{\ea}{\end{eqnarray}}
\newcommand{\bs}{\begin{subequations}}
\newcommand{\es}{\end{subequations}}

\newcommand{\grts}{\raise.3ex\hbox{$>$\kern-.75em\lower1ex\hbox{$\sim$}}}
\newcommand{\lets}{\raise.3ex\hbox{$<$\kern-.75em\lower1ex\hbox{$\sim$}}}

\begin{document}
\vspace*{1cm}

\title{Could the LHC two-photon signal correspond to the heavier scalar
in two-Higgs-doublet models?}

\author{P.\ M.\ Ferreira}\thanks{E-mail: ferreira@cii.fc.ul.pt}
\affiliation{Instituto Superior de Engenharia de Lisboa,
	1959-007 Lisboa, Portugal}
\affiliation{Centro de F\'{\i}sica Te\'{o}rica e Computacional,
    Faculdade de Ci\^{e}ncias,
    Universidade de Lisboa,
    Av.\ Prof.\ Gama Pinto 2,
    1649-003 Lisboa, Portugal}
\author{Rui Santos}\thanks{E-mail: rsantos@cii.fc.ul.pt}
\affiliation{Instituto Superior de Engenharia de Lisboa,
	1959-007 Lisboa, Portugal}
\affiliation{Centro de F\'{\i}sica Te\'{o}rica e Computacional,
    Faculdade de Ci\^{e}ncias,
    Universidade de Lisboa,
    Av.\ Prof.\ Gama Pinto 2,
    1649-003 Lisboa, Portugal}
\author{Marc Sher}\thanks{E-mail: mtsher@wm.edu}
\affiliation{High Energy Theory Group, College of William and Mary,
	Williamsburg, Virginia 23187, U.S.A.}
\author{Jo\~{a}o P.\ Silva}\thanks{E-mail: jpsilva@cftp.ist.utl.pt}
\affiliation{Instituto Superior de Engenharia de Lisboa,
	1959-007 Lisboa, Portugal}
\affiliation{Centro de F\'{\i}sica Te\'{o}rica de Part\'{\i}culas,
    Instituto Superior T\'{e}cnico, Technical University of Lisbon,
    1049-001 Lisboa, Portugal}

\date{\today}

\begin{abstract}
LHC has reported tantalizing hints for a
Higgs boson of mass 125 GeV decaying into two photons.
We focus on two-Higgs-doublet Models,
and study the interesting possibility that
the heavier scalar $H$ has been seen,
with the lightest scalar $h$ having thus far escaped
detection.
Non-observation
of $h$ at LEP severely constrains the
parameter-space of two-Higgs-doublet
models.
We analyze cases where the decay
$H \rightarrow h h$ is kinematically allowed,
and cases where it is not,
in the context of type I, type II,
lepton-specific, and flipped models.
\end{abstract}


\maketitle

\section{Introduction}

After a decades-long wait,
experiments at the LHC have finally started to probe the
electroweak symmetry breaking sector of the electroweak
theory.
In the Standard Model,
this entails a single scalar field whose particle
remnant is the Higgs boson.
There is no fundamental principle why the physical theory
should involve only one scalar field,
however.
As is the case for fermions,
scalars could appear in multiple families.
The simplest such extension 
is the two-Higgs-doublet model (2HDM),
recently reviewed in Ref.~\cite{ourreview},
where details and extensive references to the original
literature may be found.

Generic 2HDMs exhibit flavor-changing neutral currents,
for which there are stringent bounds arising from mixing in the
neutral meson systems,
such as $K$-$\bar{K}$ and $B_d$-$\bar{B_d}$.
The usual solution invokes a discrete $Z_2$ symmetry.
In the type I 2HDM, 
all fermions couple to a single Higgs doublet.
The lepton-specific model is similar to type I,
in that all quarks couple only to one Higgs doublet,
but the leptons couple exclusively to the other Higgs doublet.
In the type II 2HDM, up-type quarks and charged leptons
couple to one Higgs doublet,
while down-type quarks couple to the other.
The flipped model is obtained from type II by flipping
the leptons;
up-type quarks couple to one Higgs doublet,
while down-type quarks and leptons couple to the other.

LHC \cite{atlas,cms} has reported some hints
for a $125$ GeV state decaying into two photons.
In the context of 2HDMs,
this state could be the light scalar $h$,
the pseudoscalar particle $A$,
or the heavy scalar $H$. 
Consequences of the first possibility were
investigated in Ref.~\cite{fsss1};
consequences of the second were discussed
in Ref.~\cite{burdman}.
Here we focus on the third possibility:
that there is indeed a scalar particle
of 125 GeV,
but that this is the heavier of the two scalars, $H$.
This would mean that the lightest scalar $h$ should have,
thus far, evaded detection.
The combined requirements on $H$ and $h$ place
stringent limits on the parameter space.
We will consider two qualitatively distinct cases.
In case~1,
$m_h=105$ GeV and $m_H=125$ GeV,
thus precluding the decay $H \rightarrow h h$.
In case~2,
$m_h=50$ GeV and $m_H=125$ GeV,
implying that $H \rightarrow h h$ is kinematically allowed.
In both cases,
we assume that the charged scalars and the
neutral pseudoscalar are sufficiently heavy
(or their couplings sufficiently suppressed)
that they do not affect LHC data.

We follow Ref.~\cite{fsss1} and plot
the various experimental constraints in the
($\sin\alpha,\tan\beta$) plane.
Here,
$\alpha$ is the rotation angle which diagonalizes
the neutral scalar mass matrix, and the angle $\beta$ is defined as
\be
\tan{\beta} \equiv \frac{v_2}{v_1},
\ee
where $v_1$ and $v_2$ are the vacuum expectation values
of the two scalar doublets,
and $v \equiv \left( v_1^2 + v_2^2 \right)^{1/2}$ is the
Standard Model vacuum expectation value.
The two parameters $\alpha$ and $\beta$
determine the interactions of the various Higgs fields
with the vector bosons and
(given the fermion masses)
with the fermions.

In all four models,
the coupling of the neutral Higgs $h$ ($H$) to the
$W$ and $Z$ bosons is the same as in the Standard Model,
multiplied by $\sin(\beta-\alpha)$ ($\cos(\beta-\alpha)$).
The other relevant couplings are listed in
Table~\ref{tab:couplings}.
\begin{table}
\begin{center}
\begin{tabular}{|c|p{1.2in}|p{1.2in}|p{1.2in}|p{1.2 in}|}  \hline
{} & Type I  & Type II & Lepton-specific & Flipped     \\
\hline
$htt $    & $\cos\alpha/\sin\beta$   & $\cos\alpha/\sin\beta$
& $\cos\alpha/\sin\beta$ & $\cos\alpha/\sin\beta$
\\
\hline
$hbb $    & $\cos\alpha/\sin\beta$        & $-\sin\alpha/\cos\beta$
& $\cos\alpha/\sin\beta$  & $-\sin\alpha/\cos\beta$
\\
\hline
$h \tau\tau $ & $\cos\alpha/\sin\beta$     & $-\sin\alpha/\cos\beta$
& $-\sin\alpha/\cos\beta$   & $\cos\alpha/\sin\beta$
\\
\hline\
$Htt $    & $\sin\alpha/\sin\beta$     & $\sin\alpha/\sin\beta$
& $\sin\alpha/\sin\beta$ & $\sin\alpha/\sin\beta$
\\
\hline
$Hbb $    & $\sin\alpha/\sin\beta$     & $\cos\alpha/\cos\beta$
& $\sin\alpha/\sin\beta$ & $\cos\alpha/\cos\beta$
\\
\hline
$H\tau\tau $    & $\sin\alpha/\sin\beta$     & $\cos\alpha/\cos\beta$
& $\cos\alpha/\cos\beta$ & $\sin\alpha/\sin\beta$
\\
\hline
\end{tabular}
\end{center}
\caption{Yukawa couplings of $t,b,\tau$ to the neutral Higgs scalars,
$h$ and $H$,
in the four different models,
divided by the Standard Model couplings.}
\label{tab:couplings}
\end{table}
In the type I and lepton specific models,
on the one hand,
and in the type II and flipped models at
small $\tan\beta$,
on the other,
the production through gluon fusion is determined
by the coupling to the top in the triangle loop.
In contrast,
in the type II and flipped models at large $\tan\beta$,
the triangle with the bottom quark
becomes relevant and may even exceed
that with the top quark.

\begin{figure}[h!]
\centering
\hspace{-1.cm}
\includegraphics[height=2.66in,angle=0]{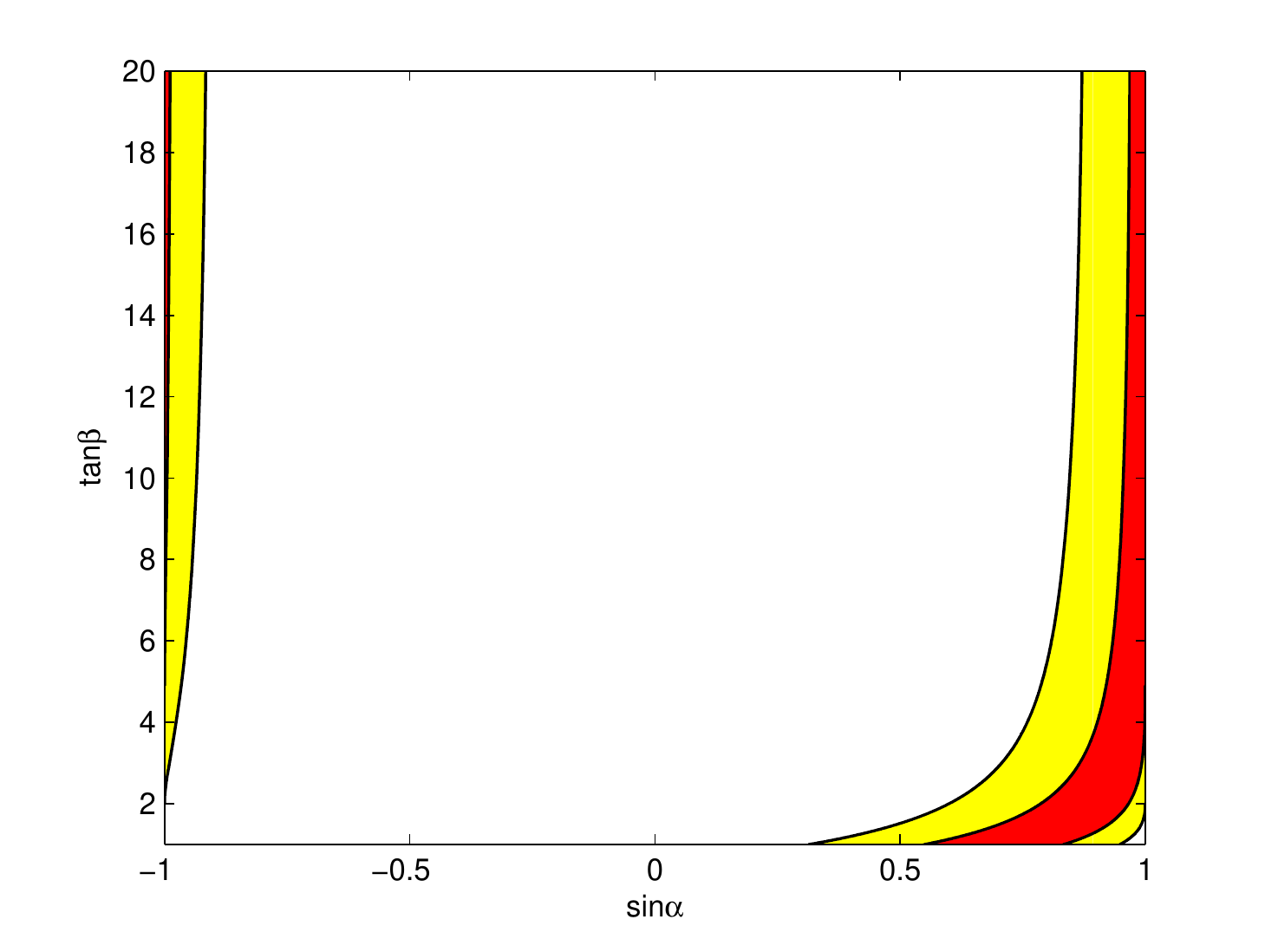}
\vspace{-0.4cm}
\caption{We plot the lines of constant $\sin^2{(\beta-\alpha)}$
in the  ($\sin\alpha,\tan\beta$) plane.
For $m_h=105$ GeV ($m_h=50$ GeV),
LEP constrains the parameters to lie within the
light yellow (dark red) shaded areas.}
\label{figsin2ab}
\end{figure}

LEP experiments searched for associated production of a
light Higgs up to masses around $115$ GeV \cite{LEP}.
In 2HDMs,
rates with $hVV$ couplings ($V = Z, W$) are suppressed by
$\sin^2{(\beta-\alpha)}$,
which the LEP data constrains to lie below $\sim 0.2$
for $m_h=105$ GeV \cite{LEP,Pedro}.
This implies a very stringent constraint on the
($\sin\alpha,\tan\beta$) plane,
shown for $m_h=105$ GeV as the light yellow shaded areas in Fig.~\ref{figsin2ab}.
For $m_h = 50$ GeV,
LEP constrains $\sin^2{(\beta-\alpha)}$ to lie
below $\sim 0.04$,
leading to even smaller allowed regions,
shown in Fig.~\ref{figsin2ab} as
dark red areas.\footnote{Strictly speaking,
the LEP constraint is even tighter
in the exact fermiophobic limit
of the type I model \cite{LEP_fermiophobic}.}

For the most part,
the LEP constraint in Fig.~\ref{figsin2ab}
forces $\sin\alpha$ to be close to $\pm 1$
and $\cos\alpha$ close to zero,
with a severe impact on the observability of the
lightest Higgs.
Fig.~\ref{figsin2ab} is easy to understand in the gaugephobic
limit $\alpha=\beta$.
In that case the hVV couplings vanish,
satisfying trivially the LEP bound.
For $\tan{\beta}=1$,
$\sin{\alpha}=1/\sqrt{2}$,
which lies on the right hand allowed region in 
Fig.~\ref{figsin2ab}.
Fig.~\ref{figsin2ab} is also easy to understand in
the large $\tan{\beta}$ limit.
In that case $|\sin{(\beta - \alpha)}| \approx |\cos{\alpha}|$,
which the LEP limit forces to lie below $\sqrt{0.2}$,
forcing $|\sin{\alpha}| > \sqrt{0.8}$,
for $m_h=105$ GeV.

In the large $\tan{\beta}$ limit,
gluon-gluon fusion through the top triangle loop
is suppressed by $\cos{\alpha}$,
but production through vector boson fusion and
associated production are also suppressed by
$\sin{(\beta - \alpha)}$.
Since the Standard Model predictions for the
latter are much suppressed with respect to the former,
we may ignore them.
At the Tevatron,
Higgs searches rely on associated production,
so the $W$ and $Z$ can provide a tag for the events. 
As a result,
the $\sin^2(\beta-\alpha)$ suppression factor eliminates
any useful Tevatron bounds on the Higgs mass (beyond the LEP bounds).

We define the number of $H$ and $h$ events relative to their
Standard Model values, respectively by:
\begin{eqnarray}
\eta_H &=& 
\frac{N^{2HDM}_H}{N^{SM}},
\nonumber\\
\eta_h &=& 
\frac{N^{2HDM}_h}{N^{SM}},
\end{eqnarray}
where $N$ is the number of events,
obtained through multiplication of
the production cross section by the
relevant branching ratio ${\rm BR}$.

\section{$m_h$ above the $H \to hh$ threshold}

We begin with the type I model,
where, for $H \rightarrow \gamma \gamma$, we find
\be
\eta_H = 
\left(\frac{\sin\alpha}{\sin\beta}\right)^2
\frac{{\rm BR}^{2HDM}_H}{{\rm BR}^{SM}}.
\ee
The branching ratio in the Standard Model for
a Higgs mass of $125.0$ GeV is $0.00228$ \cite{gg}
with a 5\% error.\footnote{Naturally,
the precise position of the lines of constant $\eta_H$
in the figures presented in this article will depend on the
exact value taken for ${\rm BR}^{SM}$.
This also affects whether or not a line, for example $\eta_H=1$,
is allowed in some 2HDM.
For consistency,
we will take ${\rm BR}^{SM}$ directly from our programs.
We are interested mainly in the qualitative features
to be probed in the foreseeable future.
Detailed higher order simulations will only be
relevant once the precision of the
experiments increases dramatically.}

\begin{figure}[htb]
\centering
\hspace{-1.cm}
\includegraphics[height=2.66in,angle=0]{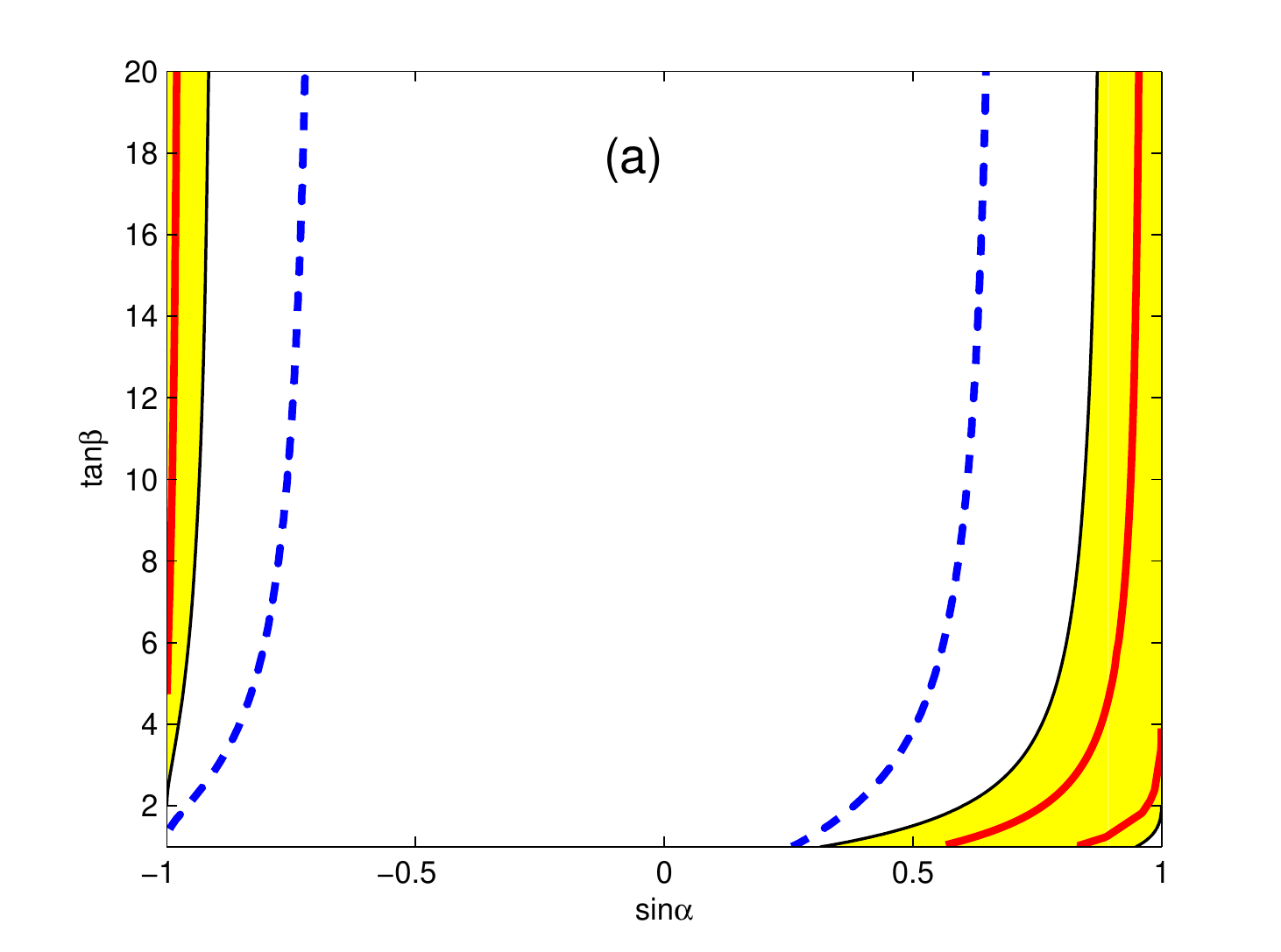}
\includegraphics[height=2.66in,angle=0]{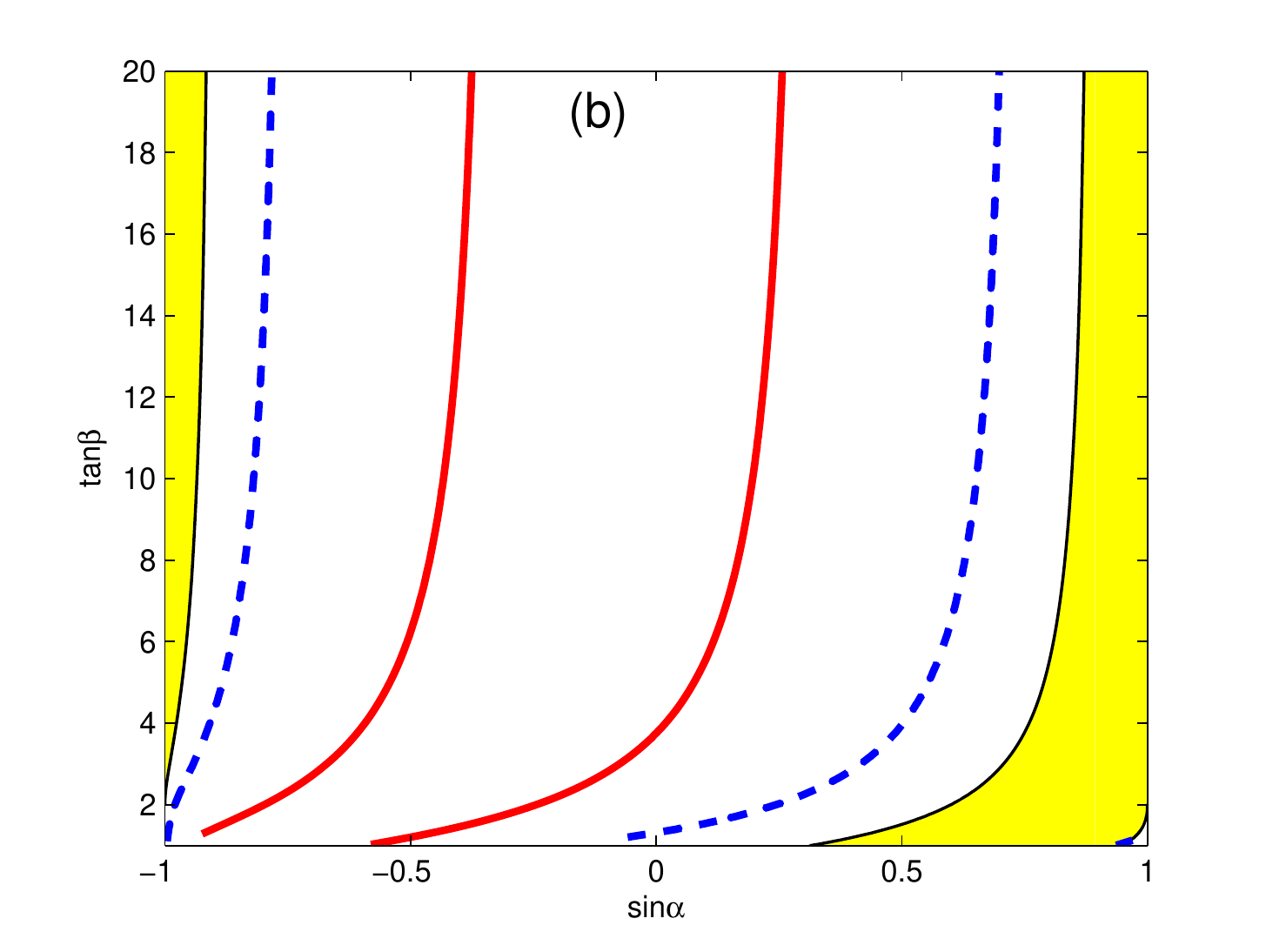}
\vspace{-0.4cm}
\caption{For type I 2HDM,
we plot the lines of equal ratios $\eta_H=N^{2HDM}_H/N^{SM}$ (a) and
$\eta_h =N^{2HDM}_h/N^{SM}$ (b)
in the ($\sin\alpha,\tan\beta$)
plane for the $H,h \rightarrow \gamma\gamma$ signal.
Along the red (solid) lines the ratio is $1$,
and along the blue (dashed) lines it is $1/2$.
The maximum $\eta_H$ in the allowed region lies around
1.4.}
\label{HhggI}
\end{figure}

In Fig.~\ref{HhggI}a,
we plot the $\eta_H=1/2$
and $\eta_H = 1$ lines in the
($\sin\alpha,\tan\beta$) plane.
Notice that it is difficult to exceed the Standard Model
value in the context of type I models.
We see that the LEP constraint on the light Higgs
forces the $\gamma\gamma$ decay of the heavy Higgs to lie
very close to its SM value.
For example,
$\eta_H=1/2$ is excluded.
This is consistent with its detectability
in the $\gamma \gamma$ channel at the LHC.   
For $h \rightarrow \gamma \gamma$,
we find
\be
\eta_h = 
\left(\frac{\cos\alpha}{\sin\beta}\right)^2
\frac{{\rm BR}^{2HDM}_h}{{\rm BR}^{SM}},
\ee
which is plotted in Fig.~\ref{HhggI}b for
$\eta_h=1/2$ and $\eta_h=1$.
All these values are excluded,
meaning that,
for this scenario,
the lightest Higgs decay into $\gamma \gamma$
will not be seen at LHC in the near future.
As in $\gamma \gamma$,
we find that $H \rightarrow VV$
might be seen at rates comparable to the SM,
while $h \rightarrow VV$ cannot.

\begin{figure}[h!]
\centering
\hspace{-1.cm}
\includegraphics[height=2.66in,angle=0]{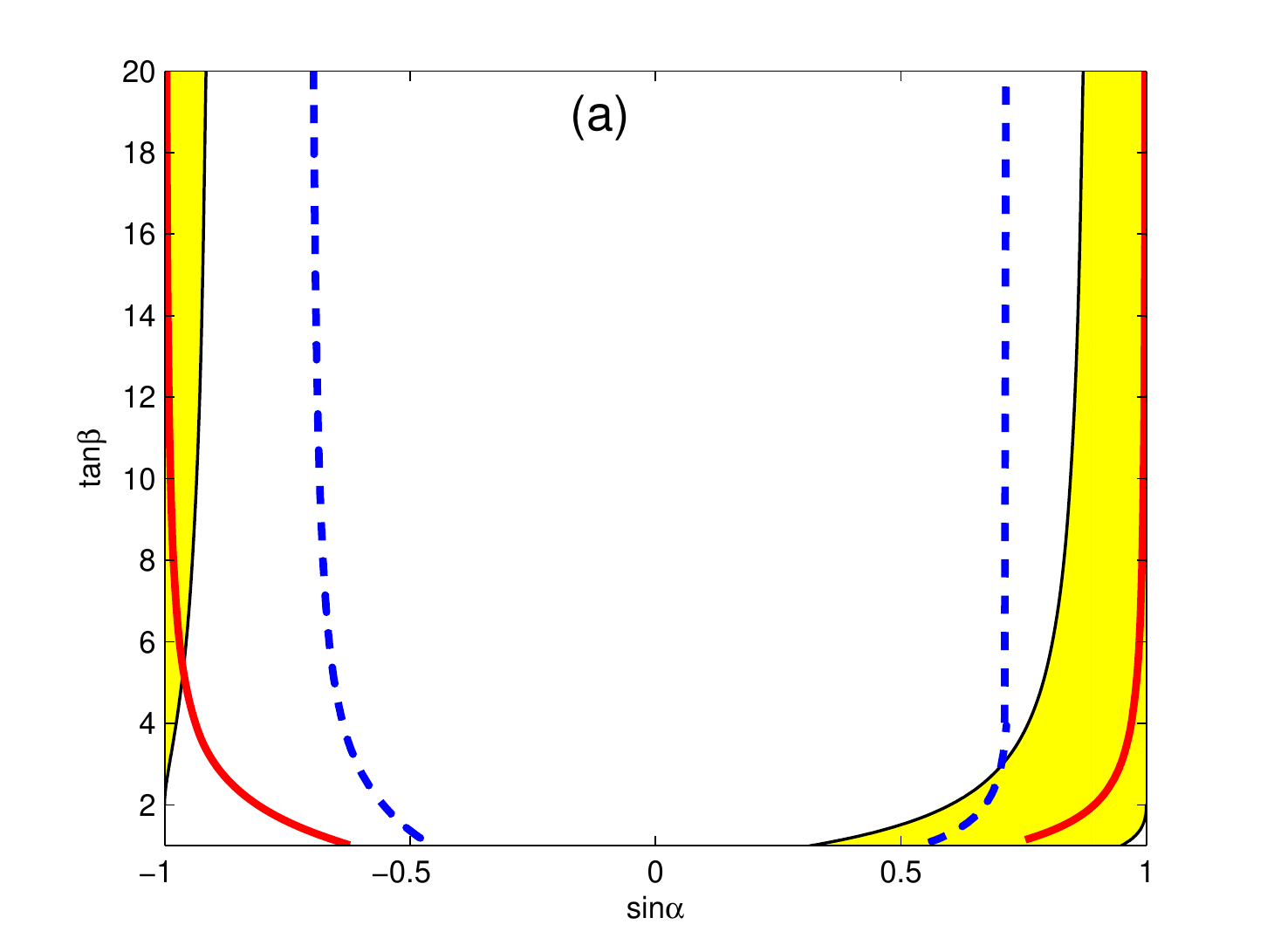}
\includegraphics[height=2.66in,angle=0]{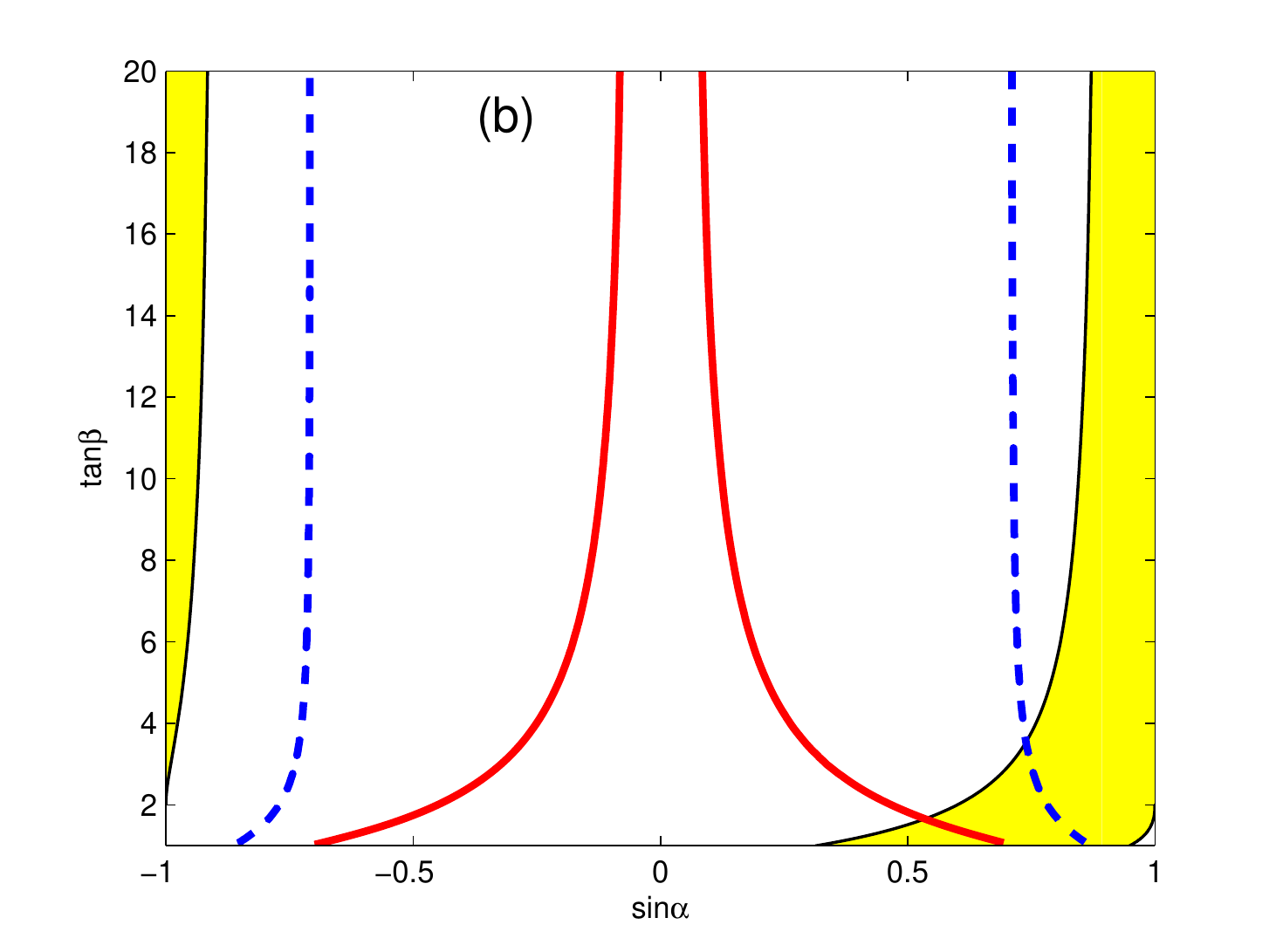}
\vspace{-0.4cm}
\caption{For type I 2HDM,
we plot the lines of equal ratios $\eta_H=N^{2HDM}_H/N^{SM}$ (a) and
$\eta_h =N^{2HDM}_h/N^{SM}$ (b)
in the ($\sin\alpha,\tan\beta$)
plane for the $H,h \rightarrow b \bar{b}$ signal.
Along the red (solid) lines the ratio is $1$,
and along the blue (dashed) lines it is $1/2$.
In the type I model,
$H,h \rightarrow \tau^+ \tau^-$
exhibit very similar features.
}
\label{HhbbI}
\end{figure}
An interesting situation for type I 2HDM arises 
in the decays into $b \bar{b}$,
shown in Fig.~\ref{HhbbI}. 
We find that $H$ can decay into $b \bar{b}$,
with $\eta_H = 1$ or with $\eta_H = 1/2$,
in a small region close to $(\sin{\alpha},\tan{\beta})=(0.7,2)$.
This is the same region in which $h \rightarrow b \bar{b}$
could have the SM rate.
The same conclusions hold for
$H \rightarrow \tau^+ \tau^-$
and $h \rightarrow \tau^+ \tau^-$,
respectively.
This raises the interesting possibility that
the decays into $b \bar{b}$ and $\tau^+ \tau^-$
could be sensitive to \textit{both}
the heavy and the light Higgs scalars,
while only $H$ can be seen in $\gamma \gamma$ and $VV$.

\begin{figure}[htb]
\centering
\hspace{-1.cm}
\includegraphics[height=2.66in,angle=0]{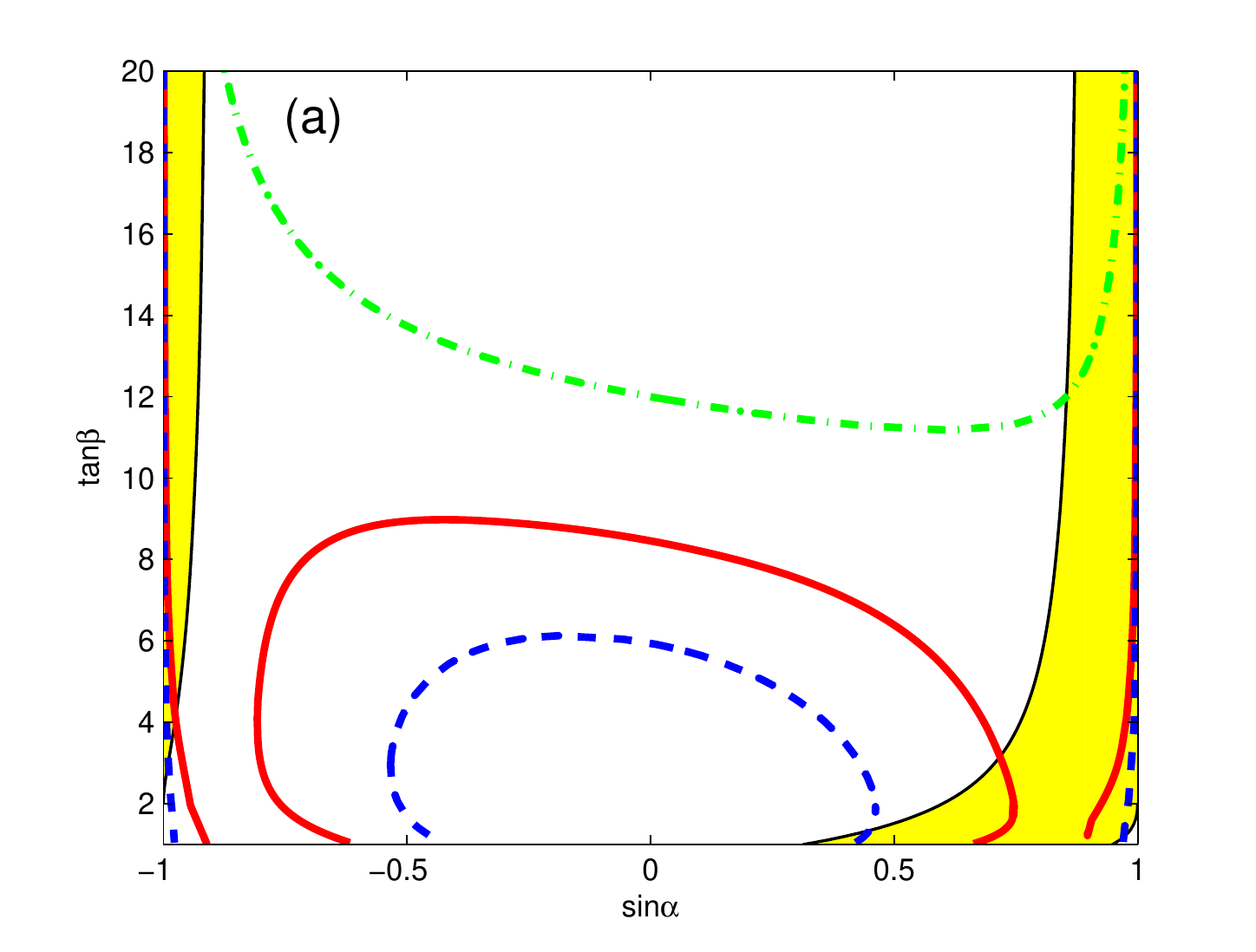}
\includegraphics[height=2.66in,angle=0]{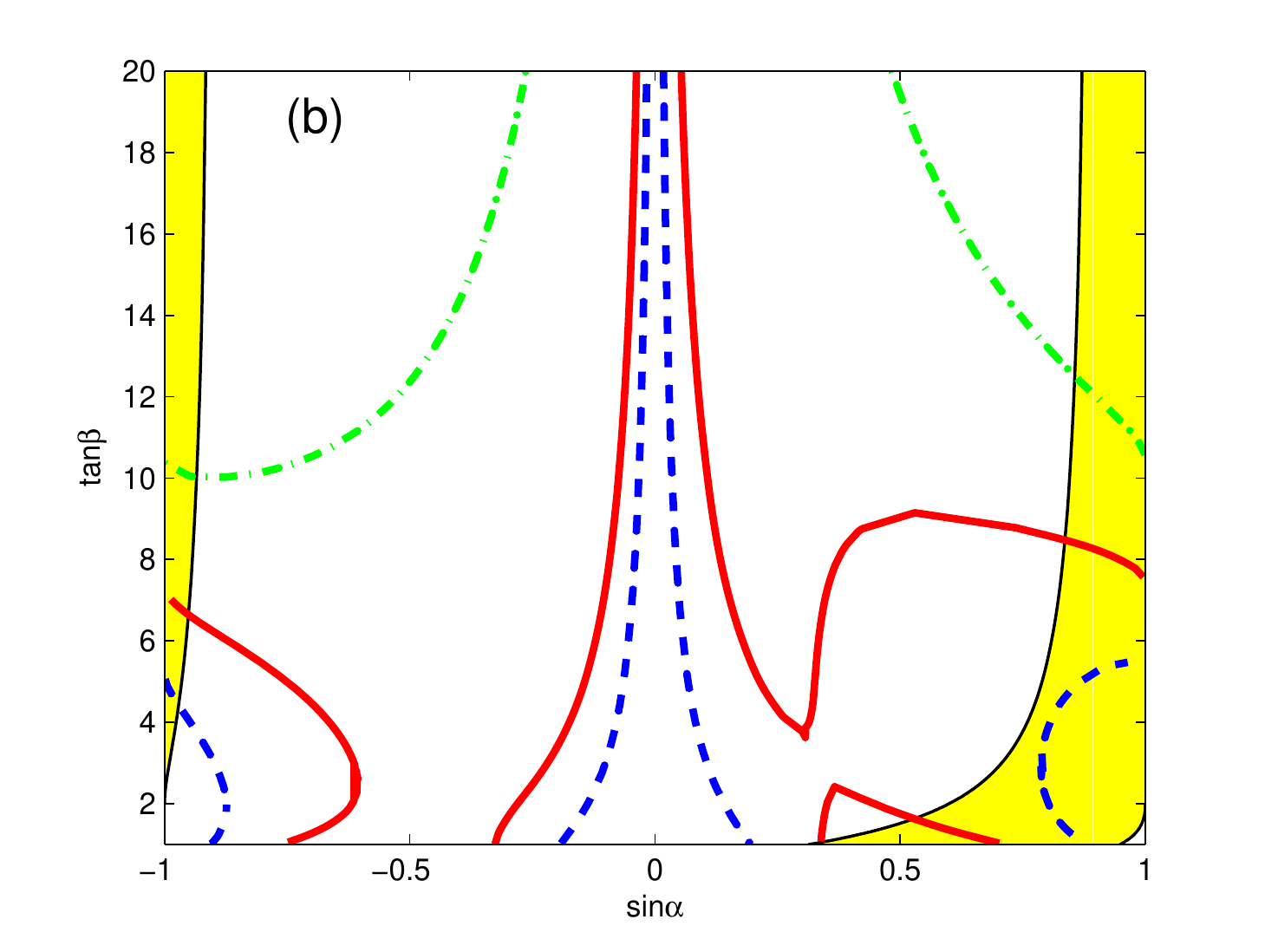}
\vspace{-0.4cm}
\caption{For type II 2HDM,
we plot the lines of equal ratios $\eta_H=N^{2HDM}_H/N^{SM}$ (a) and
$\eta_h =N^{2HDM}_h/N^{SM}$ (b)
in the ($\sin\alpha,\tan\beta$)
plane for the $H,h \rightarrow b \bar{b}$ signal.
Along the red (solid) lines, the ratio is $1$,
along the blue (dashed) lines, it is $1/2$,
and along the green (dash-dotted) lines, it is $2$.
}
\label{HhbbII}
\end{figure}

A priori,
the type II model has a different behavior,
especially at large $\tan\beta$,
due to the enhancement of the bottom quark Yukawa coupling,
affecting both the production and decay of the Higgs.
The production cross section for $g g \rightarrow h$
was calculated with HIGLU \cite{Spira:1995mt}.
Here the decay $H \rightarrow \gamma \gamma$ can have
$\eta_H=1/2$, $1$ or even $2$ in the regions
of Fig.~\ref{figsin2ab} consistent with the LEP constraint on $h$,
but, again,
$h \rightarrow \gamma \gamma$ is undetectable.
Similar conclusions occur for decays into $VV$.
But the situation may improve with respect to the type I model,
concerning $b \bar{b}$,
as shown in Fig.~\ref{HhbbII}. 
We see that both $H \rightarrow b \bar{b}$ and $h \rightarrow b \bar{b}$
could occur at rates \textit{twice the SM rate},
for $\sin{\alpha} > 0.8$ and $\tan{\beta} >  13$.
The same behavior is seen in $\tau^+ \tau^-$.

\begin{figure}[htb]
\centering
\hspace{-1.cm}
\includegraphics[height=2.66in,angle=0]{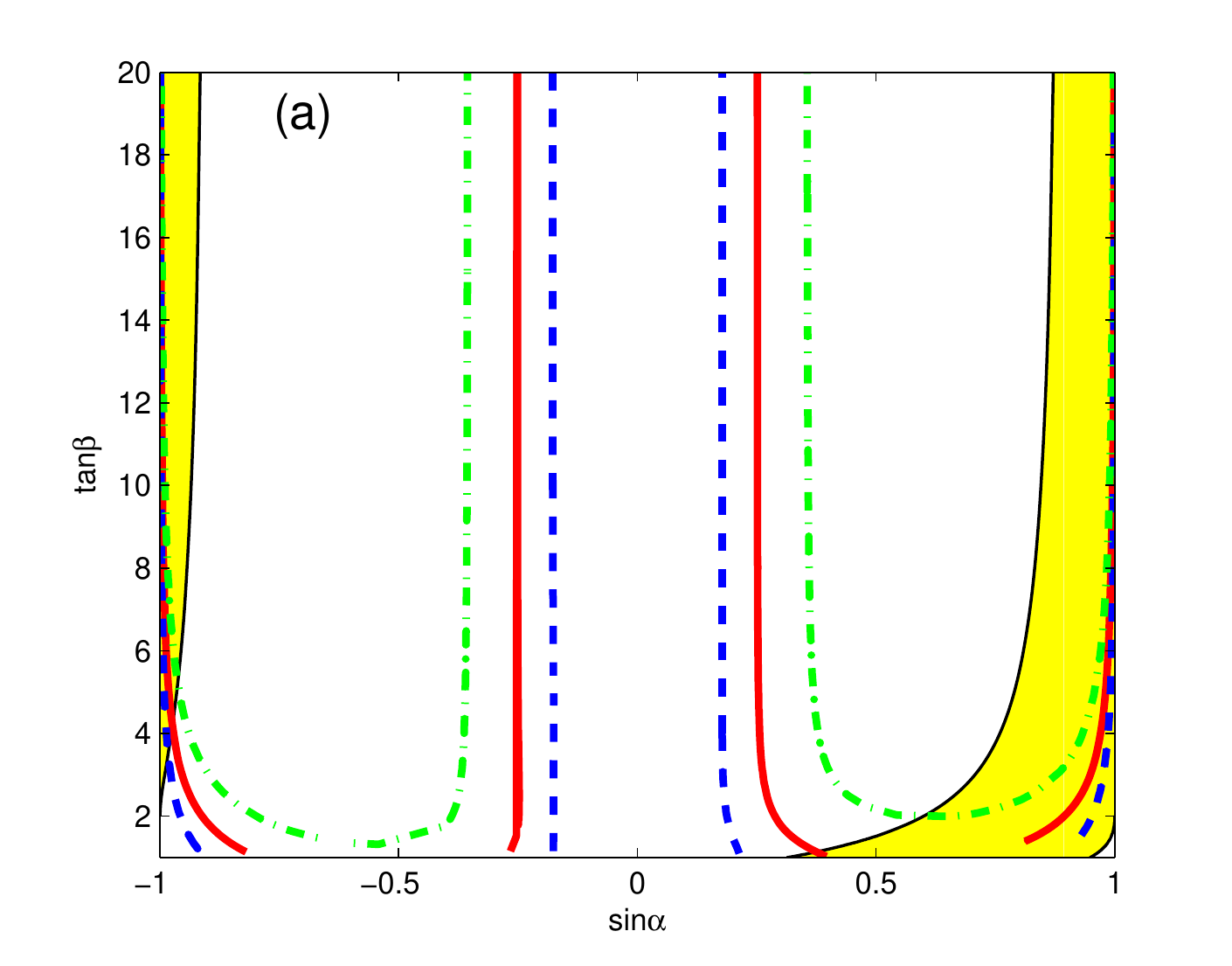}
\includegraphics[height=2.66in,angle=0]{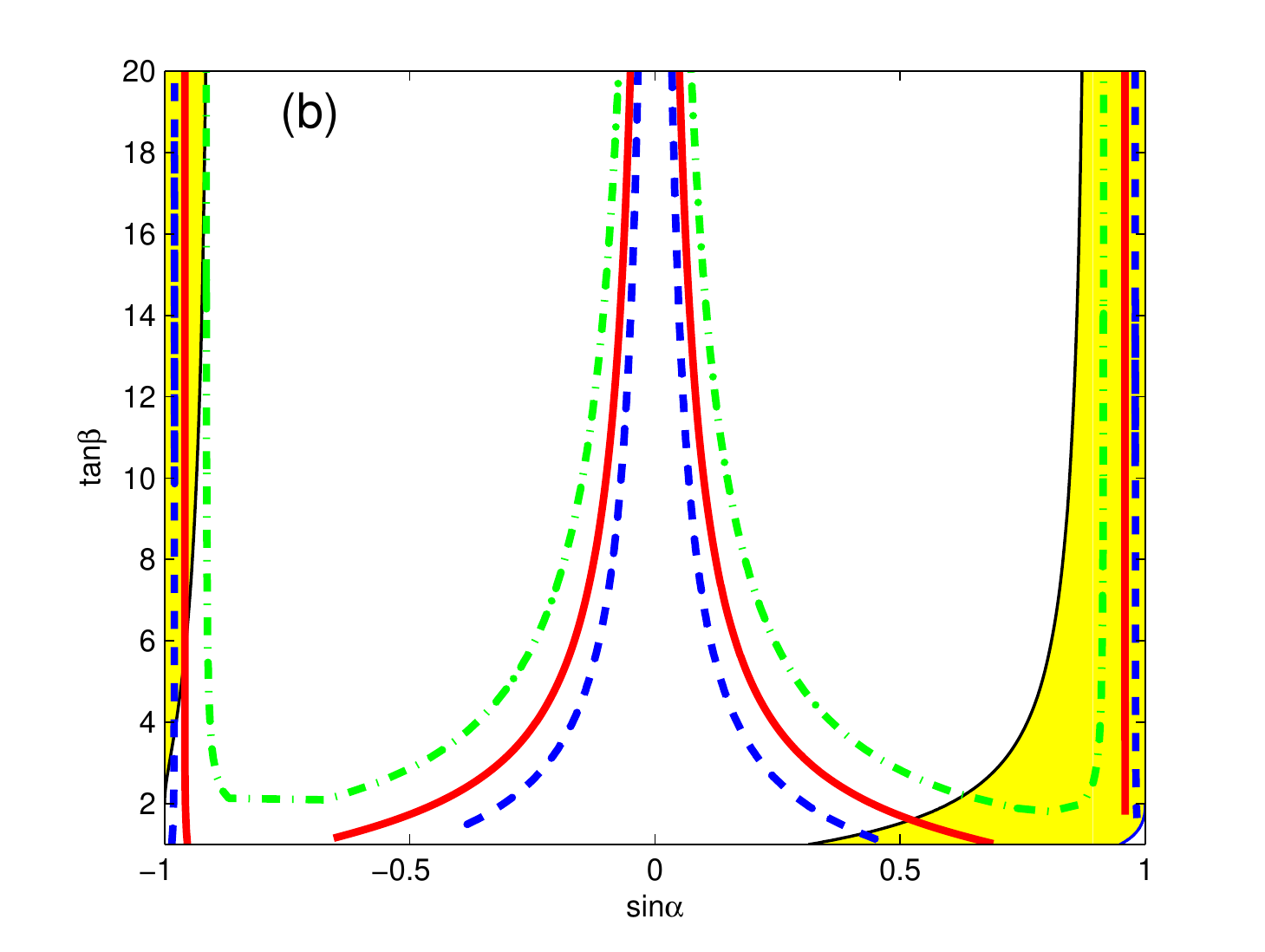}
\vspace{-0.4cm}
\caption{For lepton specific 2HDM,
we plot the lines of equal ratios $\eta_H=N^{2HDM}_H/N^{SM}$ (a) and
$\eta_h =N^{2HDM}_h/N^{SM}$ (b)
in the ($\sin\alpha,\tan\beta$)
plane for the $H,h \rightarrow \tau^+ \tau^-$ signal.
Along the red (solid) lines, the ratio is $1$,
along the blue (dashed) lines, it is $1/2$,
and along the green (dash-dotted) lines, it is $2$.
}
\label{HhtautauIII}
\end{figure}

Next we consider the lepton-specific model.
As in the type I model,
$h \rightarrow \gamma \gamma$ is unobservably
small,
while $H \rightarrow \gamma \gamma$ may be detected.
There is a subtle difference:
improving the precision of the
$H \rightarrow \gamma \gamma$ measurement
will imply a smaller region in
the $(\sin{\alpha},\tan{\beta})$ plane
for the lepton-specific model than implied
for the type I model.
The same holds for $VV$.
The decays into $b \bar{b}$
have features similar to those for
model I, shown in Fig.~\ref{HhbbI}.
In particular,
detection of $H \rightarrow b \bar{b}$ at SM rates
is possible for large $\sin{\alpha}$ and any value
for $\tan{\beta}$,
but simultaneous detection of
$h \rightarrow b \bar{b}$ around SM rates is only
possible for low values of $\tan{\beta}$.
Unlike model I,
here the situation for decays into $\tau^+ \tau^-$
is very different from $b \bar{b}$,
as shown in Fig.~\ref{HhtautauIII}. 
We see that the decay into $h \rightarrow \tau^+ \tau^-$ could be
substantially larger than in the SM,
a point made in Refs.~\cite{fsss1,Ack2}.
Here we stress that such a large
enhancement could occur for
\textit{both} $H \rightarrow \tau^+ \tau^-$
and $h \rightarrow \tau^+ \tau^-$.
The rate for $H \rightarrow \tau^+ \tau^-$ ($h \rightarrow \tau^+ \tau^-$)
could even reach about ten (three) times the SM
in the LEP allowed region.
Thus, ATLAS and CMS are already starting to place upper
limits on $\tan{\beta}$ for a given $\alpha$ in this model
\cite{tautau}.

We now turn to the flipped model.
Its $\gamma \gamma$ and $VV$ decay features follow
those of the type II model.
For the $b \bar{b}$ events,
we also re-obtain the type II model features,
shown in Fig.~\ref{HhbbII},
while the decays into $\tau^+ \tau^-$
follow instead Fig.~\ref{HhbbI},
common to the $b \bar{b}$  and $\tau^+ \tau^-$
channels in the type I model.

Finally,
we notice that none of the conclusions on this
section hinge on the presence or absence
of a term in the Higgs potential
softly breaking  the discrete $Z_2$ symmetry.

\section{$m_h$ below the $H \to hh$ threshold}

When $H \rightarrow h h$ is kinematically allowed
(such as in our second test case, $m_H = 125$ GeV and $m_h=50$ GeV),
we must consider the triple vertex \cite{dubinin,kanemura}
\be
\lambda_{Hhh}
\propto
\frac{\cos{(\beta - \alpha)}}{\sin{(2 \beta)}}
(m_H^2 + 2 m_h^2) \sin{(2 \alpha)}
\left[ 1 -
x \left(\frac{3}{\sin{(2 \beta)}}-\frac{1}{\sin{(2 \alpha)}}\right)
\right],
\label{LHhh}
\ee
where
\be
x = \frac{2 \mu_{12}^2}{m_H^2 + 2 m_h^2},
\ee
and $\mu^2_{12}$ allows for the inclusion in the
Higgs potential of a possible term
softly breaking the discrete $Z_2$ symmetry.

\subsection{Without soft-breaking}

We discuss the $\mu_{12}=0$ case in this section,
leaving the $\mu_{12} \neq 0$ case for the next section.
Generically,
when the $H \rightarrow h h$ channel is opened,
all other branching ratios are much suppressed and,
in particular,
$H$ could not even be seen in the $\gamma \gamma$ channel.
This violates our working hypothesis that current
LHC hints correspond indeed to $H \rightarrow \gamma \gamma$.
As a result,
we are interested in regions where $\lambda_{Hhh}$ is close
to zero.
It is easy to find such regions in the $(\sin{\alpha}, \tan{\beta})$,
when $\mu_{12}=0$.
One may have $\sin{\alpha}= -1,0,1$ or,
from $\cos{(\beta - \alpha)}=0$,
$\beta = \alpha \pm \pi/2$,
leading to $\tan{\beta} = - \sqrt{1-\sin^2{\alpha}}/\sin{\alpha}$.
Of these,
only the $\sin{\alpha} \approx \pm 1$ regions are consistent
with the $\sin^2{(\beta-\alpha)} \lessapprox 0.04$ LEP bound,
shown as the dark red regions in Fig.~\ref{figsin2ab}.
Therefore,
it is only close to $\sin{\alpha} \approx \pm 1$ that $H$
may be visible in $\gamma \gamma$,
or in any further channel other than $H \rightarrow h h$.
However,
this a necessary but not a sufficient condition,
since,
in the regions consistent with small $\lambda_{Hhh}$,
the couplings into the relevant channels
might themselves be suppressed.

\begin{figure}[htb]
\centering
\hspace{-1.cm}
\includegraphics[height=2.66in,angle=0]{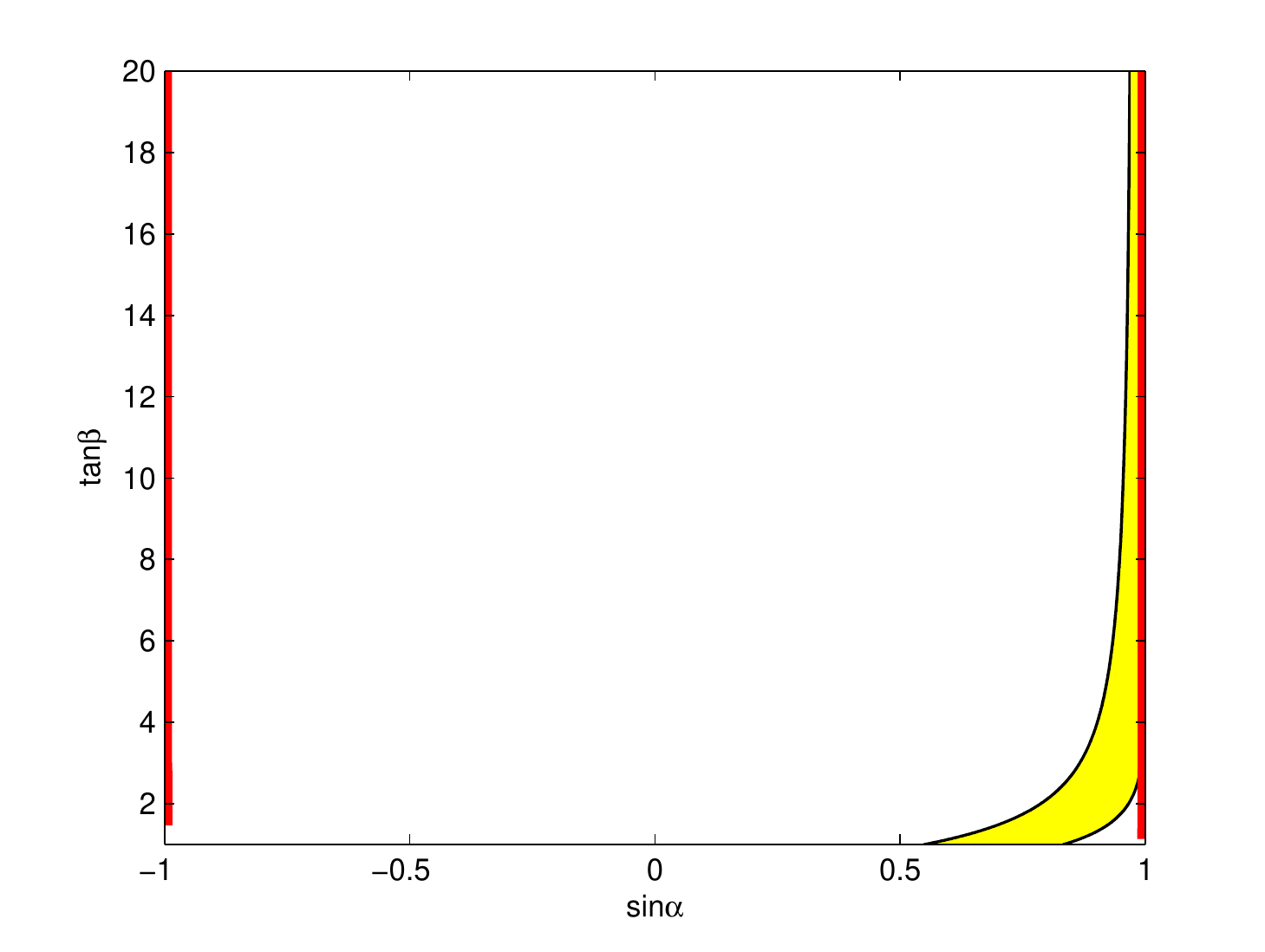}
\includegraphics[height=2.66in,angle=0]{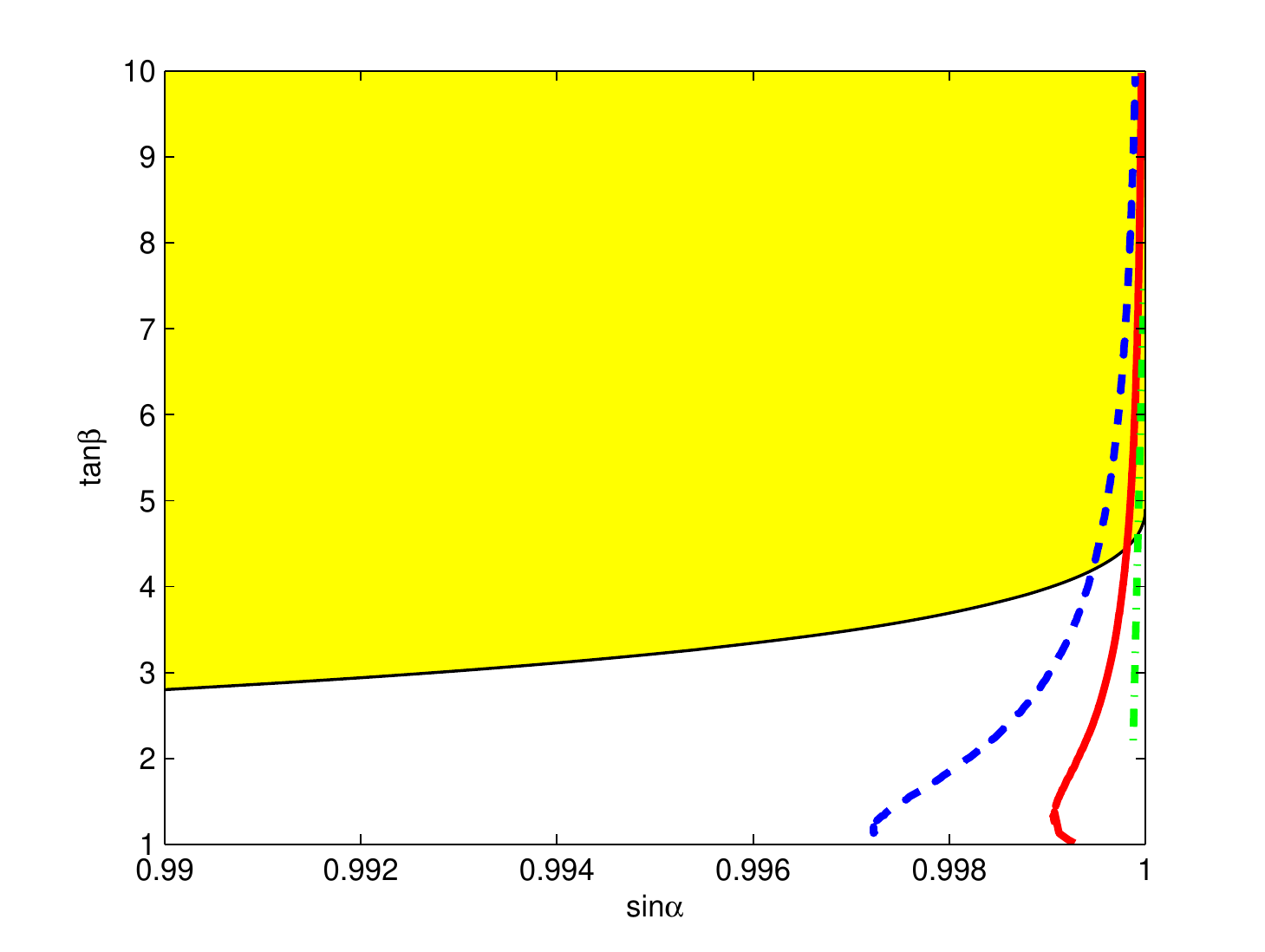}
\vspace{-0.4cm}
\caption{For type II 2HDM,
we plot the lines of equal ratios $\eta_H=N^{2HDM}_H/N^{SM}$ 
in the ($\sin\alpha,\tan\beta$)
plane for the $H \rightarrow \gamma \gamma$ signal.
With the scale shown on the left,
the lines for $\eta_H = 1/2$,
$1$ and $2$ cannot be resolved.
The leftmost allowed region is hidden by the red line.
In the plot on the right we enlarge the
region close to $\sin{\alpha}=1$,
with $\eta_H=1$ along the red (solid) line,
$\eta_H=1/2$ along the blue (dashed) line,
and $\eta_H=2$ along the green (dash-dotted) line.
Notice that the light yellow region shown here
corresponds to the dark red region in Fig.~\ref{figsin2ab}.
}
\label{modII_h50_H_phph}
\end{figure}

Fig.~\ref{modII_h50_H_phph} shows $\eta_H$ for the decay
$H \rightarrow \gamma \gamma$ for the type II and flipped models.
This means that $H \rightarrow \gamma \gamma$ may occur
at levels twice the SM,
if and only if $\sin{\alpha} \approx \pm 1$,
as predicted.
Equivalent figures hold for the type I and lepton specific
models, except that $\eta_H$ may not exceed a value around
the SM.
The results are approximately the same
for $H \rightarrow VV$.
We conclude that all models are consistent
with a $125$ GeV $H$ detected through its
$\gamma \gamma$ and $VV$ decays,
as long as $\sin{\alpha} \approx \pm 1$.

In the scenario under study in this section,
the $b \bar{b}$ and $\tau^+ \tau^-$ experiments
turn out to be crucial to discriminate among
the four models.
Indeed,
combining the LEP bound with the observability of
$H \rightarrow \gamma \gamma$,
we find the following $H$ properties:
it might be seen
in both decays,
for the type I model;
it might be seen in $b \bar{b}$ but not in
$\tau^+ \tau^-$,
for the lepton specific model;
it might be seen in $\tau^+ \tau^-$ but not in
$b \bar{b}$,
for the flipped model;
and it will not be seen in either,
for the type II model.

The $\sin{\alpha} \approx \pm 1$ constraint
also has a very strong impact on the detectability
of the light scalar $h$.
To avoid the LEP bound,
$\sin^2{(\beta-\alpha)} \lessapprox 0.04$,
and $h$ is close to gaugephobic.
Thus,
it cannot be seen in $VV$,
regardless of the specific 2HDM considered.
This also means that Higgs detection which
requires associated production of a W or Z,
as needed at the Tevatron,
will be strongly suppressed.
We have checked that $h \rightarrow \gamma \gamma$
and $h \rightarrow b \bar{b}$ is undetectable,
while $h \rightarrow \tau^+ \tau^-$ is
only detectable in the lepton specific model.
Notice that,
in the scenario $m_H=125$ GeV,
$m_h=50$ GeV, and $\mu_{12}=0$,
the lepton specific model has a very interesting
prediction:
$H$ may be seen in $\gamma \gamma$, $VV$,
and $b \bar{b}$ at rates around the SM value,
but it will not show up in $\tau^+ \tau^-$,
while $h$ exhibits exactly the opposite features.

\subsection{With soft-breaking}

\begin{figure}[htb]
\centering
\hspace{-1.cm}
\includegraphics[height=2.66in,angle=0]{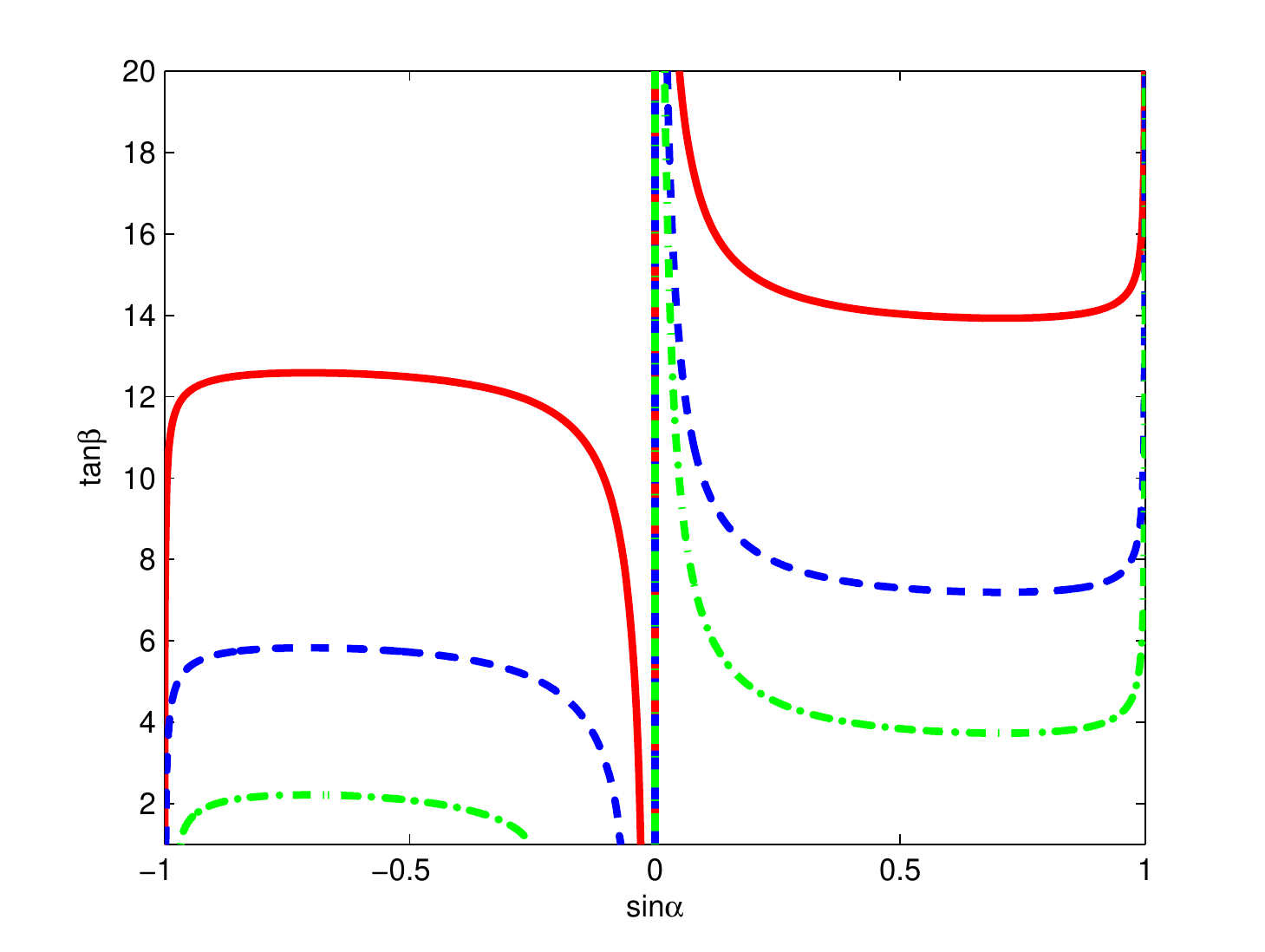}
\vspace{-0.4cm}
\caption{Lines of constant $x$ which satisfy
Eq.~\eqref{zeros_x}:
$x=0.05$ on the red (solid) line,
$x=0.1$ on the blue (dashed) line,
$x=0.2$ on the green (dash-dotted) line.
Along these lines $\lambda_{Hhh}=0$.
Notice that the light yellow region shown here
corresponds to the dark red region in Fig.~\ref{figsin2ab}.
}
\label{xx}
\end{figure}

As in the previous section,
requiring $H \rightarrow \gamma \gamma$
observability means that
we are interested in regions where $\lambda_{Hhh}$ is close
to zero.
Besides the conditions found in the previous section,
where the pre-factor in Eq.~\eqref{LHhh} vanishes,
$\lambda_{Hhh}$ will also vanish
when
\be
x = \frac{\sin{(2 \alpha)} \sin{(2 \beta)}}{3 \sin{(2 \alpha)} - \sin{(2 \beta)}}.
\label{zeros_x}
\ee
Fig.~\ref{xx} shows lines in the ($\sin\alpha,\tan\beta$)
plane where the expression between squared parenthesis
in Eq.~\eqref{LHhh} (and, thus, $\lambda_{Hhh}$)
vanishes.\footnote{Some authors use $M^2 = m_{12}^2/\sin{(2 \beta)}$ 
instead of $\mu_{12}^2$ \cite{kanemura}.
A plot of lines with constant $M^2$ will differ
from Fig.~\ref{xx},
especially for negative $\sin{\alpha}$.
Of course,
the physics is the same.}
Thus,
at least in principle and given
some chosen region in the ($\sin\alpha,\tan\beta$) plane,
there is a judicious choice of $\mu_{12}$ guaranteeing that
$H \rightarrow \gamma \gamma$ is not swamped
by $H \rightarrow h h$.
Notice that this is a necessary, but far from a sufficient
condition, for $H \rightarrow \gamma \gamma$ observability.

\begin{figure}[htb]
\centering
\hspace{-1.cm}
\includegraphics[height=2.66in,angle=0]{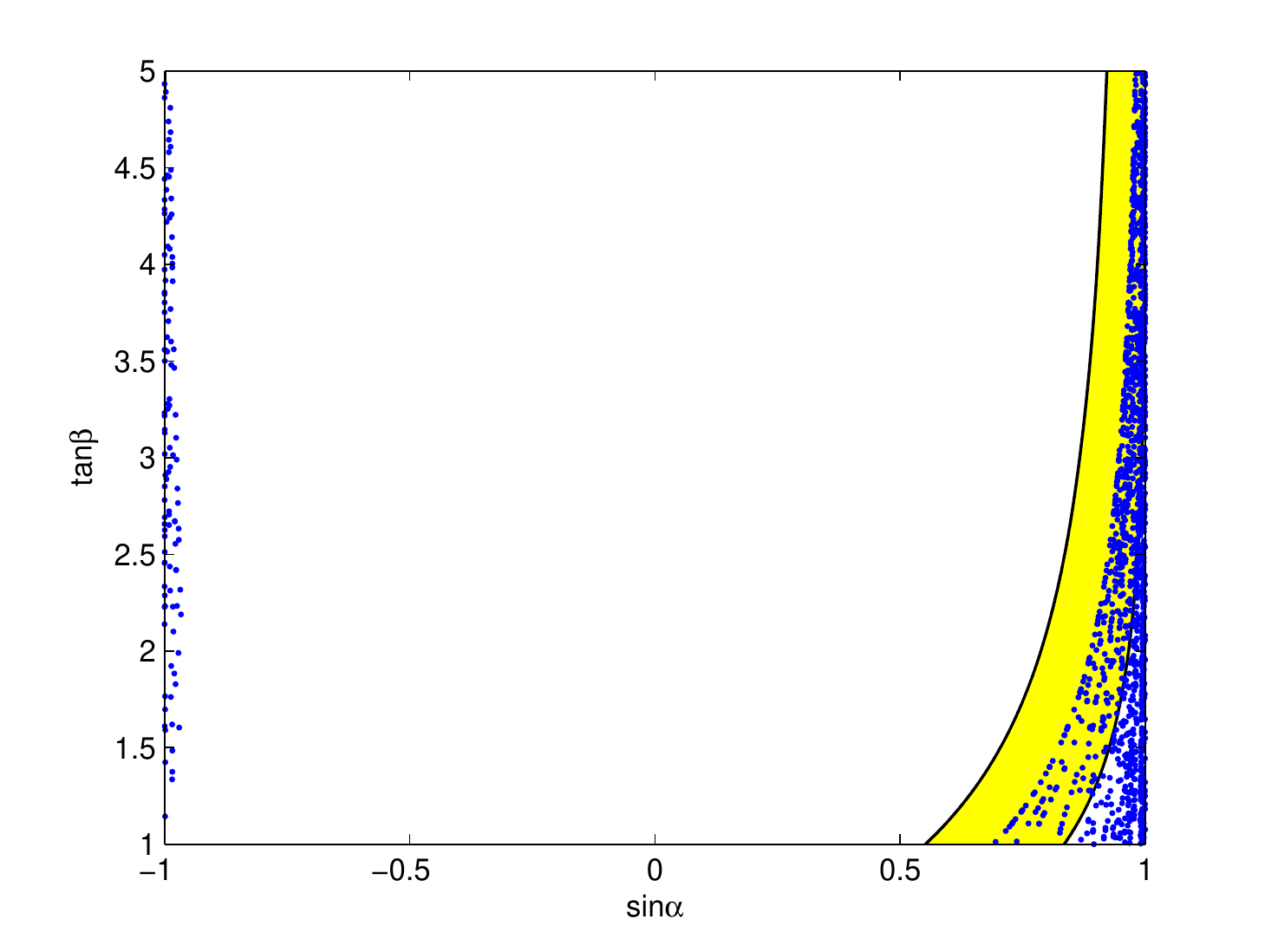}
\vspace{-0.4cm}
\caption{Scatter plot for the $H \rightarrow \gamma \gamma$ signal
in the type II 2HDM,
with $\eta_H=N^{2HDM}_H/N^{SM} = 1$.
The allowed region on the left hand side of
Fig.~\ref{figsin2ab} is not shown because here
we keep $\tan{\beta}$ below 5.
}
\label{m12neq0_modII}
\end{figure}

We conclude that,
in the presence of $\mu_{12} \neq 0$,
we might have $H \rightarrow \gamma \gamma$ at levels consistent
with LHC hints in regions away from the 
$\sin{\alpha} = \pm 1$ constraints implied by Fig.~\ref{modII_h50_H_phph}.
This is shown as a scatter plot in Fig.~\ref{m12neq0_modII},
drawn for the type II model and for random choices of $\mu_{12}$.
One can now fill almost the entire LEP allowed region.
This is even more so in the type I model.
In this case,
the phenomenology is very similar to the $m_h = 105$ GeV case.

\section{Conclusions}

At long last, the experimental exploration
of electroweak symmetry breaking has
begun.
The recent results at the LHC,
if confirmed in the next few months,
indicate that a scalar field exists at around 125 GeV
which decays into two photons at a rate which is not very
dissimilar to the Standard Model rate.
It is natural to now begin discussing the consequences
of such a field in the context of extensions to the Standard Model.

The simplest such extension is the two-Higgs doublet model,
and we study the four versions of the model with natural
flavor conservation.
We address the question of whether the state at 125 GeV,
$H$, could be the \textit{heavier} of the two neutral scalars.
For simplicity, we assume that the charged scalar and
pseudoscalar are either sufficiently heavy or sufficiently decoupled that our
results are not substantially affected.
LEP's non-observation of a lighter Higgs, $h$,
will severely constrain the parameter-space of the models, requiring
the $h$ to be nearly gaugephobic.

In all four models, we find that the decays
$h\rightarrow \gamma\gamma, WW, ZZ$ will be unobservable.
If the $h$ mass is above $62.5$ GeV,
then $H \rightarrow h h$ is kinematically inaccessible.
In the type I model, the decays of $h$ and
$H$ into $b\bar{b}$ and $\tau^+\tau^-$ can both be
observed at a rate similar to that of the Standard Model,
whereas in the type II model, these decays can
actually both occur at rates twice that of the Standard Model;
we have delineated the parameter-space in both cases.
In the lepton-specific case,
one can have a huge enhancement in the $H\rightarrow\tau^+\tau^-$ and
$h\rightarrow\tau^+\tau^-$ rates,
the former (latter) possibly being enhanced by up to a factor of ten (three).
The flipped model is similar to the type I and
type II models.

If the $h$ mass is below $62.5$ GeV,
the $H\rightarrow hh$ is kinematically allowed,
and will generally be large.
We first study the case in which there is no term softly-breaking
the $Z_2$ symmetry which suppresses flavor-changing neutral currents.
The observation of the two-photon decay at the LHC forces one into
a small region of parameter-space in which the $Hhh$ coupling is
suppressed.
Again,
the two-photon decay of the $h$ is undetectable,
and in this region of parameter-space the $b\bar{b}$
decay is also suppressed,
as is $\tau^+\tau^-$ in all but the lepton-specific model.
In the lepton-specific model, $h\rightarrow\tau^+\tau^-$ is detectable.
With the soft-breaking term,
the region of parameter-space in which the $Hhh$ coupling is suppressed
is substantially expanded,
and can cover most of the LEP-allowed region,
leading to similar results as in the heavier $h$ case.

Should the LHC detect a second Higgs below the LEP bound decaying into
$\gamma\gamma$, the two-Higgs doublet model will only be viable if the charged
Higgs and pseudoscalar are quite light,
so our assumption that they have no effect would break down.
This, of course, would lead to more interesting
phenomenology.
Note that we have not discussed the possibility that the
lighter Higgs is above the LEP bound (say at 119 GeV).
Having the two neutral scalars so close in mass would
require substantial fine tuning,
but the possibility deserves further investigation.

\begin{acknowledgments}
We are grateful to Renato Guedes for help
with HIGLU.
The work of P.M.F., R.S., and J.P.S.
is supported in part by the Portuguese
\textit{Funda\c{c}\~{a}o para a Ci\^{e}ncia e a Tecnologia} (FCT)
under contract PTDC/FIS/117951/2010 and by an
FP7 Reintegration Grant, number PERG08-GA-2010-277025.
P.M.F. and R.S. are also
partially supported by PEst-OE/FIS/UI0618/2011.
The work of M.S. is funded by the National Science Foundation grant
NSF-PHY-1068008 and by a Joseph Plumeri Award.
The work of J.P.S. is also funded by FCT through the projects
CERN/FP/109305/2009 and  U777-Plurianual,
and by the EU RTN project Marie Curie: PITN-GA-2009-237920.
\end{acknowledgments}

\end{document}